\documentclass[prl,aps, twocolumn,floatfix]{revtex4-1}
\usepackage[utf8]{inputenc}

\usepackage{natbib}
\usepackage{graphicx}
\usepackage{color}
\usepackage[tbtags]{amsmath}
\usepackage{amssymb}
\usepackage{gensymb}
\usepackage{float}

\newcommand{\be}{\begin{equation}}
\newcommand{\ee}{\end{equation}}
\newcommand{\bea}{\begin{eqnarray}}
\newcommand{\eea}{\end{eqnarray}}

\newcommand{\la}{\langle}
\newcommand{\ra}{\rangle}

\newcommand{\lp}{\left(}
\newcommand{\rp}{\right)}

\renewcommand{\Im}{{\rm Im\,}}
\renewcommand{\vec}[1]{{\boldsymbol #1}}
\renewcommand{\phi}{\varphi}
\renewcommand{\epsilon}{\varepsilon}
\def\nn{\nonumber\\}

\begin{document}

\title{Intrinsically Undamped Plasmon Modes in Narrow Electron Bands}

\author{Cyprian Lewandowski, Leonid Levitov} 

\affiliation{Massachusetts Institute of Technology, 77 Massachusetts Ave, Cambridge, MA02139, USA}

\begin{abstract}
Surface plasmons in 2-dimensional electron systems with narrow Bloch bands feature an interesting regime in which Landau damping (dissipation via electron-hole pair excitation) is completely quenched. This surprising behavior is made possible by strong coupling in narrow-band systems characterized by large values of the ``fine structure'' constant $\alpha=e^2/\hbar \kappa v_{\rm F}$. Dissipation quenching occurs when dispersing plasmon modes rise above the particle-hole continuum, extending into the forbidden energy gap that is free from particle-hole excitations. The effect is predicted to be prominent in moir{\'e} graphene, where at magic twist-angle values, flat bands feature $\alpha\gg1$. The extinction of Landau damping enhances spatial optical coherence. Speckle-like interference, arising in the presence of disorder scattering, can serve as a telltale signature of undamped plasmons directly accessible in near-field imaging experiments.
\end{abstract}

\maketitle

\begin{figure}[!htb]
    \centering
    \includegraphics[width=0.93\linewidth]{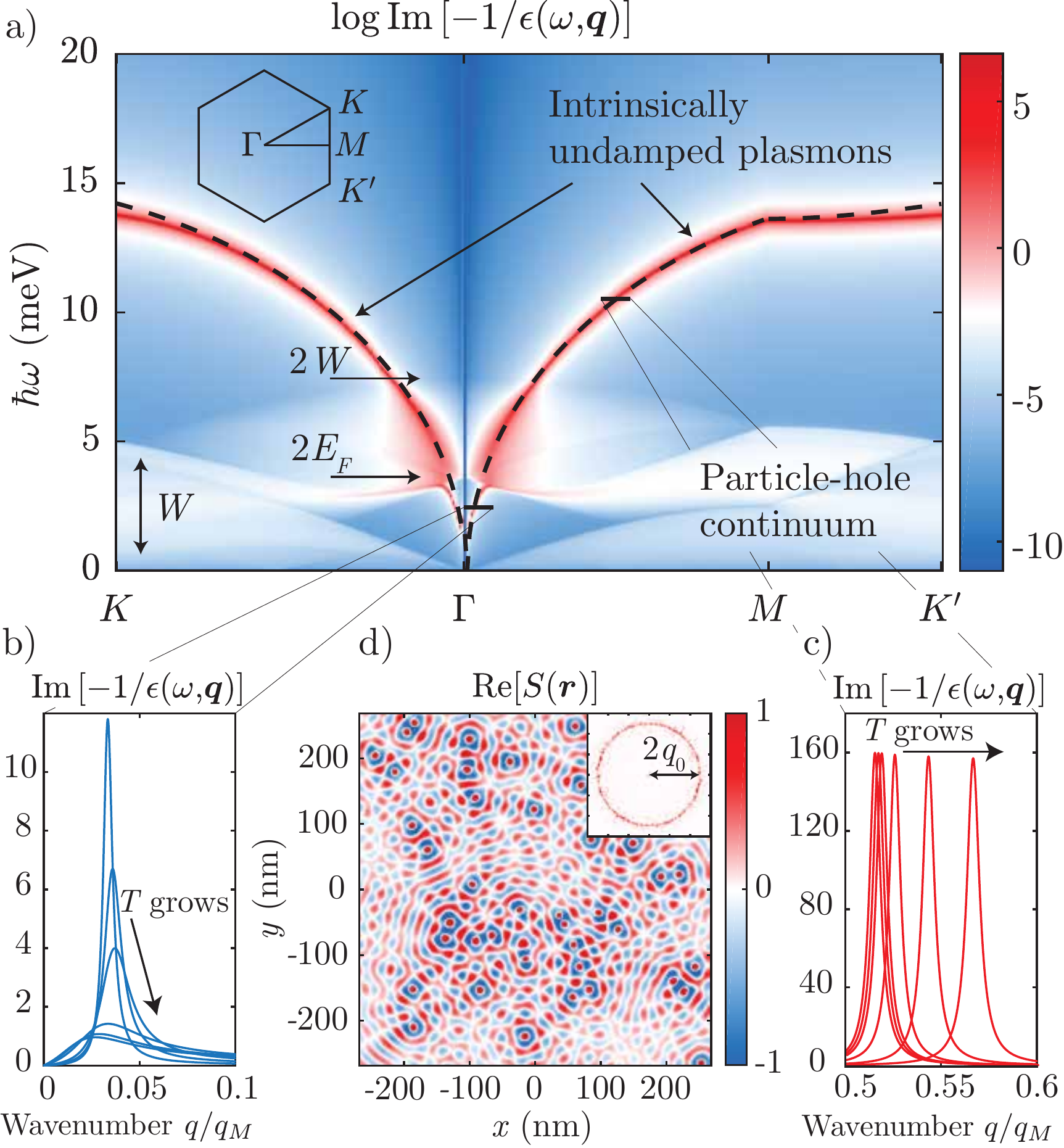}
    \caption{
    (a) Electron loss function $\Im(-1/\epsilon(\omega,\vec{q}))$ for a narrow-band toy model (the hexagonal tight-binding model) [Eq.  
      \eqref{eq:dielectric_function_tb_step_2}]. 
    Parameter values are chosen to mimic TBG bands (bandwidth $W = 3.75$ meV, lattice periodicity $L_M=13.4$ nm, Fermi energy in the conduction band at $E_{\rm F}\approx 1.81$ meV); 
    %LL: let's quote $\kappa$ in the text: and dielectric constant $\kappa = 3.03$); 
    log scale is used to clarify the relation between different features. Arrows mark the interband p-h continuum edges. 
    %The dashed line is a fit to 
    Plasmon dispersion (red line) is fitted with $\omega_{\rm p}(q) = \sqrt{\beta_q q}$ [Eq. \eqref{eq:omega=sqrt(Bq)}]  %,  \eqref{eq:beta_0} 
    %  \eqref{eq:beta_q_powers}
    %Eqs.    \eqref{eq:plasmon_dispersion_tb},  \eqref{eq:plasmon_beta_tb} 
    (dashed line).
   % (b)-(c) 
   The difference between Landau-damped (b) and undamped behavior (c) is illustrated by line cuts of plasmon resonances at the locations marked in (a), taken at %several different 
    temperatures %values 
    $T/E_{\rm F} = 0, 0.075, 0.1, 0.2, 0.3, 0.4$. Resonances broaden with $T$ in (b) and are $T$ independent in (c) 
%   Linecuts of plasmon resonances at the locations marked in panel (a), taken at %several different 
%   temperatures %values 
%    $T/E_{\rm F} = 0, 0.075, 0.1, 0.2, 0.3, 0.4$, illustrate resonance broadening due to Landau damping (b) and the 
%$T$-independent undamped behavior (c) 
(the residual resonance width models extrinsic damping due to phonons and disorder \cite{langer2010, abajo2014, %yan2012, 
ni2018, woessner2014}). % are modelled by a residual $T$-independent resonance width].
    %. The temperature-independent resonance width is due to extrinsic effects \cite{Langer_2010,doi:10.1021/ph400147y,10.1038/nphoton.2013.103, nnano.2012.59, 10.1038/s41586-018-0136-9, 10.1038/nmat4169, PhysRevB.88.121405, 10.1038/nphoton.2013.57} (phonon emission or disorder). 
    Resonances at the 3 lowest $T$ values in (c) are slightly offset for clarity. (d) Speckle pattern in scanning near-field microscopy signal  \cite{basov2012,koppens2012} $S(\vec{r})$ [Eq. \eqref{eq:correlation_function}] due to %intrinsically 
    undamped plasmons; %a corresponding square of its Fourier transform 
%    and its 
 optical coherence is manifest in Fourier spectrum $|S_\vec{k}|^2$ (inset). %We set the 
    Results shown are for plasmon momentum $q_0 = q_{M}/2\approx 0.14$ ${\rm nm}^{-1}$, where $q_{M}$  is the distance between points  $M$ and $\Gamma$, %\addLL{[let's use length units in nm. Use conventional notation for hexagonal lattice, $\Gamma$, K, K',M]}\addCL{[Done]}
    %\addCL{[Like this?]} %and took the disorder $\eta(\vec{r})$ 
    and disorder is modeled as $40$ randomly placed point defects.}
   \label{fig:fig_1}
  \vspace{-7mm}
\end{figure}

\begin{figure}[!t]
    \centering
    \includegraphics[width=1\linewidth]{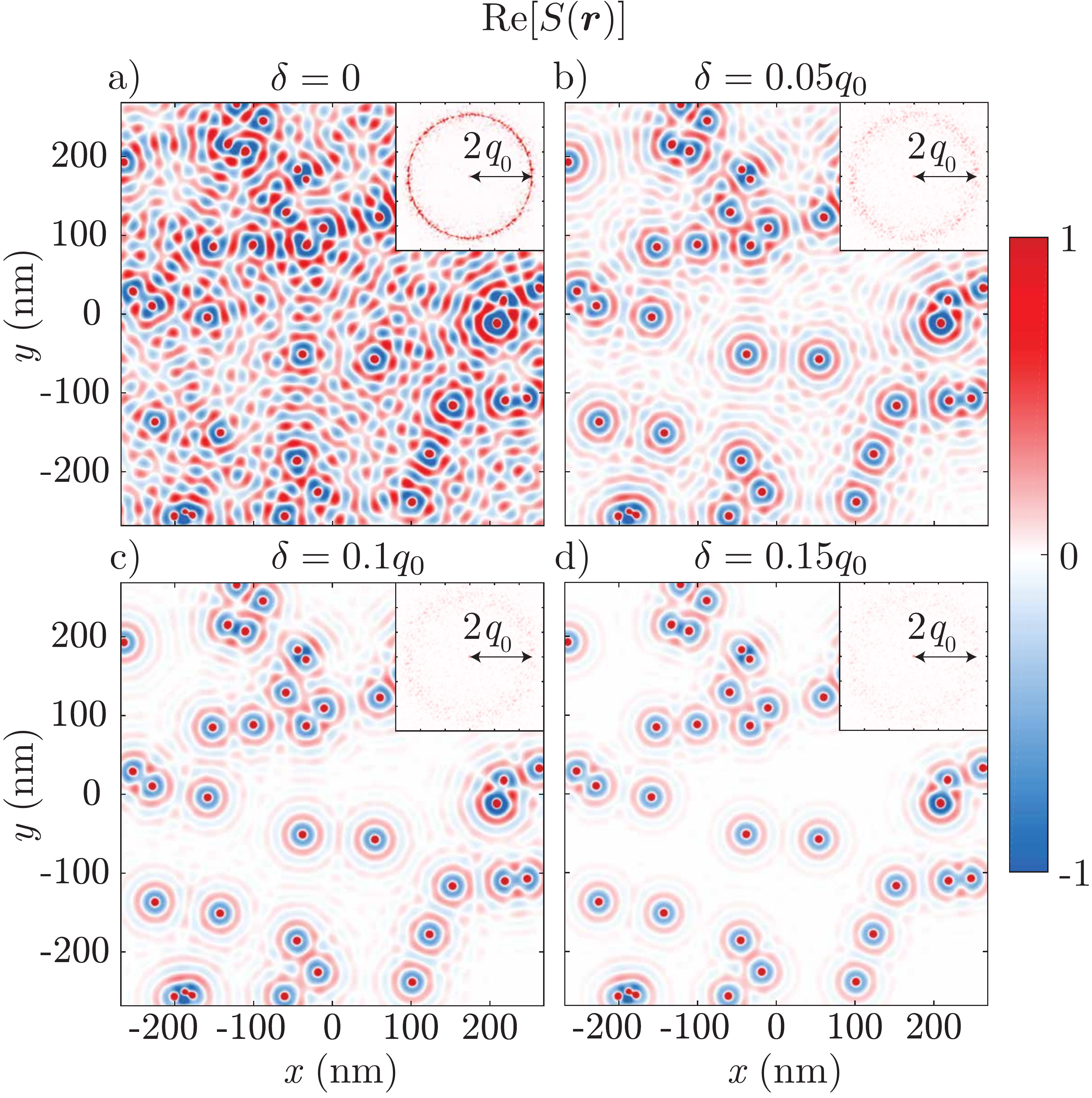}
    \caption{(a-d) Speckle patterns arising due to optical coherence of undamped plasmons in scanning near-field microscopy signal $S(\vec{r})$ [Eq. \eqref{eq:correlation_function}] at %various ratios of the 
    various ratios of the incoherent to coherent 
    damping $\delta/q_0$. Insets show the corresponding square of the speckle pattern's Fourier transform amplitude $|S_{\vec{k}}|^2$. In all panels, for clarity of comparison, we set the plasmon momentum as in Fig.  \ref{fig:fig_1}d ($q_0 = q_{M}/2 \approx 0.14$ ${\rm nm}^{-1}$) and vary only the ratio $\delta/q_0$. The disorder is taken as $40$ randomly placed Dirac delta functions.%\addCL{[CL: increase font size, change $\delta_0$ in the Fig. ]}%\addLL{[let's make length measured in nm]}
    \label{fig:fig_2}
    }
  \vspace{-5mm}
\end{figure}

Landau damping, a process by which collective mode decays into electron-hole pairs, is often taken to be an integral attribute of graphene plasmon excitations  \cite{wunsch,hwang,jablan,%koppens2011, goncalves2016, basov2016,
koppens2012,basov2012}. Here, we predict %suppression of %Landau damping 
extinction of this dissipation mechanism in materials with narrow electron bands, such as twisted bilayer graphene (TBG)  \cite{mele2010,San-Jose2012, bistritzer2011, santos2007, fang2016}. %, koshino2018, zou2018}. % {\bf add more TBG theory: Mele, Guinea.} 
%It is usually taken for granted that a fundamental limit to plasmon lifetime is set by Landau damping, a process in which collective modes decay through electron-hole pair production \cite{Landau:1946jc, nozieres1999the, 10.1038/nphoton.2013.103, doi:10.1021/nn406627u}. Here we show how this limitation imposed by Landau damping can be circumvented by using materials with narrow bands such as twisted bilayer graphene (TBG) to generate Landau damping-free collective modes even at room temperatures. 
Intrinsically undamped plasmons in narrow-band materials arise due to large fine structure parameter values $\alpha=e^2/\hbar \kappa v_{\rm F} $: strong interactions push plasmon dispersion into the energy gap above the particle-hole (p-h) continuum as illustrated in Fig.  \ref{fig:fig_1}. In this region, plasmons become 
decoupled from p-h pair excitations. Dissipation quenching, which is a surprising manifestation of strong coupling physics, %coupling physics, 
is a robust effect that persists up to room temperature and 
is insensitive to disorder (Figs.   \ref{fig:fig_1} and  \ref{fig:fig_2}). 
%This unusual behavior of plasmon modes, arising in the srong coupling regime,  renders them protected from relaxation through electron-hole pair production, a process responsible for Landau damping. 
Collective charge modes, which are damping free, are of keen interest for quantum information science as a vehicle to realize dissipationless photon-matter coupling, high-Q resonators, single-photon phase shifters and other missing components for photon-based quantum information processing toolbox  \cite{gullans2013}. 
%While %this intrinsic protection 
%the effect of 
%``lifting the curse of Landau damping'' 
%suppression 
Although extinction of Landau damping is a general effect present in all narrow electron bands, %applies to any narrow-band system, 
our analysis will focus on TBG flat bands, a system of high current interest \cite{cao2018_1, cao2018_2, yankowitz2019, stauber2016, hu2017}, in which undamped plasmons can be directly probed. %realized and 

Fig.  \ref{fig:fig_1} depicts plasmon mode %dispersion 
for a narrow-band model that mimics the key features of the TBG band. Mode dispersion (red line) and its damping are of a  %exhibits 
%features a conventional behavior 
conventional form at energies less than the bandwidth,  $\omega \lesssim W$. %\addLL{[$E$ or $\omega$?]}\addCL{[DONE]} %, where it coexists with the p-h continuum. % (here $W$ is the bandwidth). \addLL{$E$ or $\omega$?} 
At lowest energies, plasmon mode is %displaced relative to 
positioned outside the p-h continuum, as expected; this suppresses the $T=0$ Landau damping, but does not protect the mode from decaying into p-h excitations through disorder scattering or from the conventional $T>0$ Landau damping  \cite{hwang, wunsch, %hill2009,
 emani2012, low2014, principi2013, polini2008, yan2013}. At higher energies, $\omega \sim 2E_{\rm F}$ (marked by arrows in Fig.  \ref{fig:fig_1}), the mode plunges into p-h continuum and is Landau-damped at $2E_{\rm F}\lesssim \omega\lesssim 2W$, even at $T=0$. However, an interesting change occurs after the mode rises above the p-h continuum. In the forbidden gap region, $\omega>2W$, it becomes  damping-free, since at these energies there are no free p-h pairs into which plasmon could decay. This behavior is manifest in the $T$ dependence of the resonances, which are washed out  with increasing temperature at $\omega \lesssim W$ but remain sharp at $\omega>W$ even at $T\sim E_{\rm F}$ (Fig.  \ref{fig:fig_1} b and c).

%\addCL{Plasmon dispersion}
As we will see, mode dispersion %the plasmon mode (red line in Fig.  \ref{fig:fig_1}a),  
has a square root form characteristic of 2-dimensional (2D) plasmons  \cite{stern1967, principi2009},
\begin{equation}\label{eq:omega=sqrt(Bq)}
\omega_{\rm p}(q)=\sqrt{\beta_q q}
,
\end{equation} 
with a weak $q$ dependence in $\beta_q$ [Eq. \eqref{eq:beta_q_powers}].
This expression, however, is valid not just at low energies, $0 < \omega \lesssim W$,
%, where it is Landau-damped, but also even 
but also at higher energies, $\omega \gg W$, where the mode is undamped. While the dispersion in Eq. \eqref{eq:omega=sqrt(Bq)} is of the %same form as the 
conventional 2D plasmon form, we emphasize that here it takes on a different role, as it describes the plasmon mode at frequencies much higher than the carrier bandwidth, extending to
\begin{equation}
\omega_{\rm p}\sim\sqrt{\alpha}W\gg W
,\quad 
\alpha \sim 20-30
,
\label{eq:omega=sqrt(A)W}
\end{equation} 
where the high-$\alpha$ values correspond to flat bands in magic-angle moir\'e graphene. Also, unlike the conventional plasmons, the dispersion in Eq. \eqref{eq:omega=sqrt(Bq)} is not limited to longest wavelengths. Indeed, as illustrated Fig.  \ref{fig:fig_1}a, it extends to fairly high wavenumbers on the order of the mini Brillouin zone size.

The wavelengths of these plasmons are only 2 to 3 times greater than the moir\'e superlattice period. Such short wavelengths are of considerable interest for plasmonics and are within resolution of the state-of-the-art scanning near-field microscopy techniques  \cite{basov2012,koppens2012} (currently as good as $10$ nm  \cite{cohen2014,duan2012}). In addition to measuring plasmon dispersion, these techniques can be used to directly visualize the qualitative change in the damping character and strength. % (and of its strength), 
Enhanced optical coherence will manifest itself in striking speckle-like interference as illustrated in Figs.  \ref{fig:fig_1}d and  \ref{fig:fig_2}.

Indeed, because of the absence of Landau damping at the energies of interest, $\omega>W$, and also because these energies are smaller than carbon optical phonon energies, the dominant dissipation mechanism is likely to be elastic scattering by disorder. At low energies, where plasmon mode coexists with p-h continuum, disorder scattering merely assists Landau damping, allowing plasmons to decay into p-h pairs by passing some of their momentum to the lattice. However, at the energies above p-h continuum, $\omega>W$, since the 
decay into pairs is quenched, disorder will lead to predominantly elastic scattering among plasmon excitations. Such scattering preserves optical coherence and is expected to produce speckle patterns in spatial near-field images as illustrated in Fig.  \ref{fig:fig_1}d. 

To model this behavior we consider the signal $S(\vec{r})$, excited by the scanning tip and measured at the same location. % measured by the scanning tip after it has 
 Monochromatic plasmon excitation at energy $E$ %of momentum $q_0$ at position $\vec{r}$, which then 
is scattered by impurities or defects and on returning to the tip, produces signal
\be
S(\vec{r}) = J_0 \int d^2 \vec{r}' G_E(\vec{r}-\vec{r'}) \eta(\vec{r'}) G_E(\vec{r}'-\vec{r})
,
\label{eq:correlation_function}
\ee
where $\eta(\vec{r})$ is the disorder potential, $J_0$ is excitation amplitude, and 
% at position $\vec{r}'$ and 
$G_E(\vec{r})$ is the Green's function of the plasmon excitation (see supplemental information). 
%with a fixed momentum $q_0$ (see later   \eqref{eq:plasmon_green_function}\addCL{[CL: appendix?]}).
The spatial signal (Fig.   \ref{fig:fig_1}d) exhibits a characteristic speckle pattern familiar from laser physics. % \cite{dainty1975}
In graphene plasmonics, speckle-like interference provides a direct manifestation of optical coherence enhancement in the absence of Landau damping. %The square of the Fourier transform %amplitude 
Accordingly, the Fourier transform of the image, $S_{\vec k}=\int d^2r S(\vec r) e^{-i\vec k\vec r}$, yields power spectrum
$|S_{\vec{k}}|^2$ that features a ring-like structure; the ring radius is % striking circular feature at a radius 
$k=2q_0$, where $q_0$ is the plasmon excitation wavenumber (Fig.   \ref{fig:fig_1}d inset). Simple calculation, described in supplemental information, predicts power spectrum that 
sharply peaks at the ring:
\be
|S_\vec{k}|^2 \sim \frac{ |\eta_\vec{k}|^2}{|k^2-4(q_0-i\delta)^2|}
,
\label{eq:scattering_amplitude}
\ee
where $\delta$ is a parameter % in the model, discussed below, 
characterizing extrinsic damping due to phonon scattering and other inelastic processes. %the relative strength of elastic and inelastic scattering due to disorder and phonons, respectively. 
In the fully coherent regime ($\delta = 0$) the quantity $| S_\vec{k} |^2$ exhibits a power law singularity at the ring, $k=2q_0$. As the amount of incoherent scattering increases, the peak %at the ring $k=2q_0$ 
is gradually washed out. This behavior is illustrated in Fig.   \ref{fig:fig_2}.

We note %in passing 
that recent work  \cite{stauber2016}  analyzed 
%and measured  \cite{hu2017} 
interband plasmon excitations in TBG, which are 
dominated by polarization of the bands above the flat band and 
are distinct from the flat-band plasmons analyzed here. Recent experiment  \cite{hu2017} reported observation of plasmons in TBG;
however, their appeal for constructing intrinsically protected collective modes remained unnoticed in graphene %plasmonics 
literature. Also, 
%We are aware of one paper \cite{pashitskii1993} that analyzed 
plasmons in narrow bands were analyzed in the context of high-$T_c$ superconductivity  \cite{pashitskii1993}, finding that plasmon mode can rise above the flat band. However, in cuprates, unlike moir\'e graphene, the narrow band is not separated from higher bands by a forbidden energy gap, and thus, the mode studied in  ref. \cite{pashitskii1993} will plunge into a higher band before acquiring an undamped character. 

Next, we present analysis of the hexagonal-lattice toy model that mimics the key features of Landau-damped and intrinsically undamped modes in TBG. The hexagonal-lattice tight-binding model possesses the same symmetry and the same number of subbands as the flat band in TBG. We match the energy and length scales by choosing the width of a single band $W$ and the hexagonal lattice period $L_M$ identical to the parameters in TBG:  $W=3.75$ meV and $L_M=a/2\sin(\theta/2)$ is the moir\'e superlattice periodicity. For the magic angle value $\theta=1.05\degree$, using carbon spacing $a=0.246$ nm, this gives $L_M=13.4$ nm. To ensure that a unit cell of the toy model can accomodate 4 electrons just as the moir{\'e} cell does in TBG, we make the toy model 4-fold degenerate.  Comparison with plasmons for the actual TBG model, presented below, will help us to identify the features that are general as well as those which are a specific property of TBG.

Our nearest-neighbor tight-binding Hamiltonian is 
\be
H_{\rm toy}= \begin{pmatrix} 
0 &h_{\vec{k}}\\
h_{\vec{k}}^* & 0
\end{pmatrix}
,\quad 
h_{\vec{k}} = \frac{W}{3}\sum_{\vec e_j} e^{i \vec{k}\cdot\vec e_j }
,
\ee
%where the function $g_{\vec{k}}$ is the hopping element to the nearest neighbors at positions $\vec{r}_1 = (R/2, \sqrt{3} R/2)$, $\vec{r}_2 = (R/2, -\sqrt{3} R/2)$, $\vec{r}_3 = (-R,0)$. 
with the hopping matrix element $W/3$ to nearest neighbors at positions $\vec e_j=(\cos(2\pi j/3),\sin(2\pi j/3))L_M/\sqrt{3}$, 
$j=0,1,2$. Here, $W$ is the bandwidth measured from Dirac point, and the nearest neighbor distance $L_M/\sqrt{3}$ is chosen such that the lattice period of the hexagonal toy model %is exactly equal to 
matches the moir{\'e} superlattice period. %\addLL{[Why minus sign? Why do you call $R$ ``lattice period''? ]} 
Corresponding energies $E_{s,\vec{k}}$ and eigenstates $\Psi_{s,\vec{k}}$ are then
\be
E_{s,\vec{k}}=s |h_{\vec k}|
,\quad 
\Psi_{s,\vec{k}} = \frac{1}{\sqrt{2}} \begin{pmatrix} s e^{i\phi_{\vec k}} %\frac{g(\vec{k})}{|g(\vec{k})|}
\\
1
\end{pmatrix}
%, \quad
%\phi_{\vec k}={\rm arg}\,h_{\vec{k}}
,
\label{eq:tb_eigenstates}
\ee
where $\phi_{\vec k}={\rm arg}\,h_{\vec{k}}$ and the band index $s=\pm$ labels the conduction and valence band. % respectively.
%\addLL{[How about replacing $n$ with $\zeta$ or $\xi$?]}

Plasmons can be obtained from the nodes of the complex dielectric function, describing the dynamical response of a material to an outside electric perturbation: % is given by its complex dielectric function $\epsilon(\omega,\vec{q})$ defined as:
\be
\epsilon(\omega,\vec{q})=1-V_{\vec{q}}\Pi(\omega,\vec{q})
.
\label{eq:dielectric_fun_def}
\ee
Here, $V_{\vec{q}} = 2\pi e^2/\kappa q$ is the Coulomb interaction in a medium with a background dielectric constant $\kappa$, and $\Pi(\omega,\vec{q})$ is the electron polarization function. The relation in Eq.  \eqref{eq:dielectric_fun_def} is exact as long as the polarization function is defined as an exact microscopic density-density pair correlator given by a sum of all irreducible bubble diagrams. As such, this relation can yield useful information about plasmon dispersion, even when electron interactions are strong.%, which captures response of the electron system to an outside perturbation. 

Similar to the conventional analysis of plasmons in 2D systems, here a simplification occurs in the small-$q$ limit, regardless of whether the random-phase approximation (RPA) is used to evaluate $\Pi(\omega,\vec{q})$. Indeed, since the Coulomb potential diverges at small $q$, zeros of $\epsilon(\omega,\vec{q})$ are found when the polarization function is small. However, at small $q$, this quantity vanishes as $\lambda q^2/\omega^2$, a behavior that is a consequence of the general symmetry requirements (namely, gauge invariance demanding that spatially uniform external potential does not perturb density)  \cite{mahan}. This immediately yields a $q^{1/2}$ scaling for plasmon frequency at small-enough $q$.

Below, we use the RPA approach to estimate the prefactor $\lambda$ and to demonstrate that the mode $\omega\sim q^{1/2}$ extends far above the TBG p-h continuum. To compare with other systems, we recall the familiar ``classical acceleration'' behavior found for particles with parabolic dispersion: $\Pi(\omega,\vec{q}) = n q^2/ m \omega^2$, where $n$ is the charge density and $m$ is the electron band mass  \cite{mahan}. For a more general band dispersion, the ratio $n/m$ is replaced by the band Fermi energy, $\lambda\sim{E_F}/\hbar^2$  \cite{wunsch,hwang,jablan}. 
Interactions have no impact on the behavior of  $\Pi(\omega,\vec{q})$ for the parabolic band case; however, for nonparabolic bands, the band mass $m$ must change to an effective value $m^*$ described by Landau Fermi-liquid renormalization  \cite{levitov_plasmons}.

In our case, the scaling relation $\Pi(\omega,\vec{q}) \approx \lambda q^2/\omega^2$ features different values of $\lambda$ for low and high energies, $\omega \lesssim E_F$ and $\omega > 2W$. To see this, we start with the RPA expression for polarization function 
\be
\Pi(\omega,\vec{q}) = 4 \sum_{\vec{k},s,s'} \frac{ (f_{s,\vec k+\vec q}-f_{s',\vec k})F^{s s'}_{\vec{k}+\vec{q},\vec{k}}}{E_{s,\vec{k}+\vec{q}}-E_{s',\vec{k}}-\omega-i0} %\times  
%\nn
%&\times F_{n,m}(\vec{k}+\vec{q},\vec{k}) %\left\lvert\left\langle m,\vec{k}+\vec{q} | e^{i\vec{q}\cdot \vec{r}} | n, \vec{k}\right\rangle\right\rvert^2
.
\label{eq:pol_tight_binding}
\ee
Here, summation  $\sum_{\vec{k}}$ denotes integration over the Brillouin zone (BZ), the indices $s$,$s'$ run over the electron bands %or, as will be the case for TBG, also the internal degrees of freedom (spin and valley)
and the factor of $4$ in front of the summation accounts for the 4-fold degeneracy of the toy model. Here, $f_{s,\vec k}$ is the equilibrium distribution $1/(e^{\beta (E_{s,\vec k}-E_{\rm F})}+1)$, and $F^{ss'}_{\vec{k}+\vec{q},\vec{k}}$ 
%\addLL{[$\mu$ or $E_{\rm F}$?]} 
describes band coherence factors. For our toy model, 
\begin{align}
F^{ss'}_{\vec{k}+\vec{q},\vec{k}} = | \la\Psi_{s,\vec{k}+\vec{q}} |\Psi_{s',\vec{k}} \ra |^2
=\frac{ 1+ss'\cos(\phi_{\vec k+\vec q}-\phi_{\vec k})}2
\label{eq:coherence_factor_tb}
,
\end{align}
where $\Psi_{s,\vec{k}}$ are pseudospinors given in Eq.  \eqref{eq:tb_eigenstates}.

As we now show, an analytic expression for plasmon dispersion can be obtained, describing both % identical for both 
the Landau-damped and the undamped cases in a unified way. %To that end, we rewrite 
We first rewrite 
%In order to obtain an analytic approximation for the plasmon dispersion $\omega_{\rm p}$, we rewrite first 
%the expression in 
 Eq.  \eqref{eq:pol_tight_binding} %in a slightly different form 
by performing a standard replacement $\vec{k}+\vec{q}\to-\vec{k}$ in the term containing $f_{s,\vec{k}+\vec{q}}$ followed by $-\vec{k}-\vec{q}, -\vec{k} \to \vec{k}+\vec{q}, \vec{k}$ justified by the $\vec{k}\to -\vec{k}$ time-reversal symmetry. This gives
\be
\Pi(\omega,\vec q)=8 \sum_{\vec{k},s,s'} f_{s',\vec k}\frac{F^{ss'}_{\vec{k},\vec{k}+\vec{q}}(E_{s',\vec{k}}-E_{s,\vec{k}+\vec{q}})}{(E_{s,\vec{k}+\vec{q}}-E_{s',\vec{k}})^2-(\omega+i0^{+})^2}
.
\label{eq:dielectric_function_tb_step_2}
\ee
The behavior of this expression at small $\vec q$, which will be of interest for us, can be found in a closed form. %We first note that at small $\vec q$ 
In the small-$q$ limit the coherence factors behave as
\be
F^{s=s'}_{\vec{k}+\vec{q},\vec{k}} \approx 1
,\quad
F^{s=-s'}_{\vec{k}+\vec{q},\vec{k}} \approx \frac{1}{4} \lp\vec q\cdot\nabla_{\vec k}\phi_{\vec k}\rp^2\label{eq:small_q_interband}
\ee
%Namely, the coherence factor values are 
The values $O(1)$ for intraband transitions and  $O(q^2)$ for interband transitions %. This 
might suggest that the polarization function is dominated by the intraband transitions. However, as we now show, the interband and intraband contributions are of the same order of magnitude. 

Indeed, the intraband contributions, $s=s'$, can be rewritten by noting that, upon integration over $\vec{k}$, only the %$\vec k/-\vec k$ symmetric 
even-$\vec k$ part of series expansion $E_{s,\vec{k}+\vec{q}}-E_{s,\vec{k}}$ survives, giving $\Pi_1(\omega,\vec q) \approx \frac{4}{\omega^2} \sum_{\vec{k},s} f_{s,\vec k}\lp E_{s,\vec{k}+\vec{q}}+E_{s,\vec{k}-\vec{q}}-2E_{s,\vec{k}}\rp$. Expanding in small $q$, we have
\begin{align}
\Pi_1(\omega,\vec q) 
%&\approx \frac{4}{\omega^2} \sum_{\vec{k},s} f_{s,\vec k}\lp E_{s,\vec{k}+\vec{q}}+E_{s,\vec{k}-\vec{q}}-2E_{s,\vec{k}}\rp
%\\
%&
\approx \frac{4}{\omega^2} \sum_{\vec{k},s} f_{s,\vec k} (\vec q\cdot\nabla_{\vec k})^2 E_{s,\vec{k}}\label{eq:intraband_pol_approximation}
\end{align}
As a sanity check, for parabolic band %dispersion 
$E_{\vec{k}}=k^2/2m$ this yields the familiar ``classical acceleration'' result $\Pi(\omega,\vec q)=\frac{n q^2}{m \omega^2}$. 

The interband contributions, $s=-s'$, can be simplified by noting that $E_{s,\vec{k}+\vec{q}}\approx -E_{s',\vec{k}}$, giving
\begin{align}
\Pi_2(\omega,\vec q) \approx 4\sum_{\vec{k},s} f_{s,\vec k}  \frac{E_{s,\vec{k}} \lp\vec q\cdot\nabla_{\vec k}\phi_{\vec k}\rp^2}{4E_{s,\vec{k}}^2-(\omega+i0)^2}
.
\label{eq:interband_pol_approximation}
\end{align}
As a sanity check, at $T=0$ the imaginary part of $\Pi_2$, describing interband transitions, is nonzero only for $2E_{\rm F}<\omega<2W$, as expected. The real part of  $\Pi_2$ is negative at small $\omega$ and positive at large $\omega$ because the valence band contribution dominates over that of the conduction band.

Plasmon dispersion $\omega_{\rm p}$ is given by the solution of the equation $\epsilon(\omega,\vec{q})=0$ with $\Pi=\Pi_1+\Pi_2$.
Comparing the $\omega$ dependence of $\Pi_1$ and $\Pi_2$, we see that at small frequencies, $\omega< 2E_{\rm F}$, the intraband contribution $\Pi_1$ dominates. This gives the dispersion in Eq.  \eqref{eq:omega=sqrt(Bq)} with 
\be
\beta_q=\beta_0+\beta_1 q+O(q^2)\label{eq:beta_q_powers}
\ee
where the leading term $\beta_0 = 4 \alpha v_{\rm F} E_{\rm F}/\hbar$ originates from $\Pi_1$ (see supplemental information), and the subleading $q$-dependent contribution is due to $\Pi_2$. Negative sign of $\Pi_2$ translates into $\beta_1<0$, softening the dispersion at low frequencies. This behavior, which holds the limit $\omega< 2E_{\rm F}$,  agrees with refs. \cite{wunsch,hwang,principi2009}.%,jablan}.%goncalves2016}. 

In the same manner, we can obtain the dispersion 
%The new behavior that arises 
at high frequencies, $\omega> 2W$ (the intrinsically undamped regime). The analysis is again simplified by noting that, since $\alpha=e^2/\hbar \kappa v_{\rm F}\gg 1$, the relevant values of $q$ are small compared to the Brillouin zone size, and thus, the small-$q$ limit considered above is sufficient to describe this behavior. Taking both the intraband and interband contributions in the asymptotic form $\Pi_1=\lambda_1  q^2/\omega^2$, $\Pi_2=\lambda_2 q^2/\omega^2$ where $\lambda_1 \approx 2 E_{\rm F} /\hbar^2 \pi, \lambda_2 \approx 2(W-E_{\rm F})/\hbar^2\pi$ (see supplemental information), yields Eq.  \eqref{eq:omega=sqrt(Bq)} with $\beta=\frac{2\pi e^2}{\kappa}(\lambda_1+\lambda_2)$. The first term is identical to $\beta_0$ found at low frequencies, the second term is of a positive sign, $\lambda_2>0$, describing stiffening of the plasmon dispersion due to interband transitions. 

In the undamped regime, plasmon frequency peaks at $q$ values on the order of Brillouin zone scale. The peak value of $\omega_{\rm p}$, given in Eq.   \eqref{eq:omega=sqrt(A)W}, can be found by estimating the energy differences $E_{s,\vec{k}+\vec{q}}-E_{s',\vec{k}}$ in Eq.  \eqref{eq:dielectric_function_tb_step_2} as $W$ and noting that the coherence band factor for large $q$ is in general non-vanishing and of order $1$. This gives, for the practically interesting case of $E_{\rm F}\sim W$, the result  $\omega_{\rm p} \sim \sqrt{\alpha} W$, which agrees with the dispersion $\omega_{\rm p} = \sqrt{\beta q} = 2\sqrt{\alpha v_{\rm F} W q/\hbar}$ provided that $\hbar v_{\rm F} q$ saturates at $W$. Indeed,  the estimated values of $\beta_0$,$\beta$ compared with the fitted curve in Fig.  \ref{fig:fig_1}a (see supplemental information) indicate that $\omega_{\rm p}=\sqrt{\beta_q q}$ relation from Eq. \eqref{eq:omega=sqrt(Bq)} is a good approximation for the plasmon dispersion at both small and large $q$.

%\addCL{Details of the dielectric function result plotted in Fig.  \ref{fig:fig_1}a}
The dielectric function of the 2-band toy model faithfully reproduces all of the qualitative features expected for the TBG bandstructure. However, we find that, despite matching the bandwidth $W$ and lattice period to those of TBG, the resulting %  exactly, corresponding 
plasmon dispersion extends to much higher energies then those that will be found below for the actual TBG bandstructure. This is simply because the 2-band model does not account for the effects of interband polarization of higher electron bands, which renormalize the dielectric constant down and soften the plasmon dispersion. We account for this in the toy model case by rescaling the effective fine structure constant such that the resulting plasmon dispersion is comparable in magnitude with the TBG result. Specifically, in Fig.   \ref{fig:fig_1}a, we use an effective background dielectric constant $\kappa = 12.12$, which is 4 times larger than the %experimentally relevant 
dielectric constant $\kappa = 3.03$ corresponding to an air/TBG/hexagonal boron nitride (hBN) heterostructure.

%\addCL{Twisted bilayer graphene - details of the model}
%With the above understanding of narrow-band plasmonics, we can now consider the experimentally relevant case of 

Next, we turn to the analysis of plasmons in TBG flat bands at an experimentally relevant magic angle value $\theta=1.05^{\circ}$ \cite{cao2018_1,cao2018_2,yankowitz2019}. To accurately describe the TBG band structure and eigenstates, we %choose to 
use the effective continuum Hamiltonian $H_{{\mathrm TBG}}$ introduced in ref. \cite{koshino2018}. %We relegate 
The full discussion of the band structure details %calculation to 
can be found in the supplemental material; here, we only discuss 2 relevant energy scales: flat-band bandwidth $W$ and the gap $\Delta$ between the flat bands and the rest of the band structure. %While technically 
With regard to $W$ value, we note that, technically, the bandwidth of the flat-bands, as predicted by the continuum mode $H_{{\mathrm TBG}}$, is %indeed 
on the order of $W \approx 3.75$ meV. However, the %actual 
bandwidth scale relevant for the interband and intraband excitations %we estimate 
is actually closer to $\tilde{W} \approx 2$ meV, because most of the states in the band lie below %this energy 
$2$ meV. In addition, since the states with energies outside $-2\,{\rm meV}<E<2\,{\rm meV}$ are small $k$, their contribution to polarization function [Eqs.     \eqref{eq:intraband_pol_approximation} and   \eqref{eq:interband_pol_approximation}), evaluated at small $q$, is small. 
%transitions between them unlike the toy model case, is strongly suppressed in the polarization function evaluation due to the $q^2$ dependence seen in   \eqref{eq:interband_pol_approximation},   \eqref{eq:intraband_pol_approximation}. 
We also note that, while the bandgap as predicted by the continuum model is $\Delta \approx 11.75$ meV, the actual gap is still a subject of debate \cite{tomarken2019}.

\begin{figure}
    \centering
    \includegraphics[width=\linewidth]{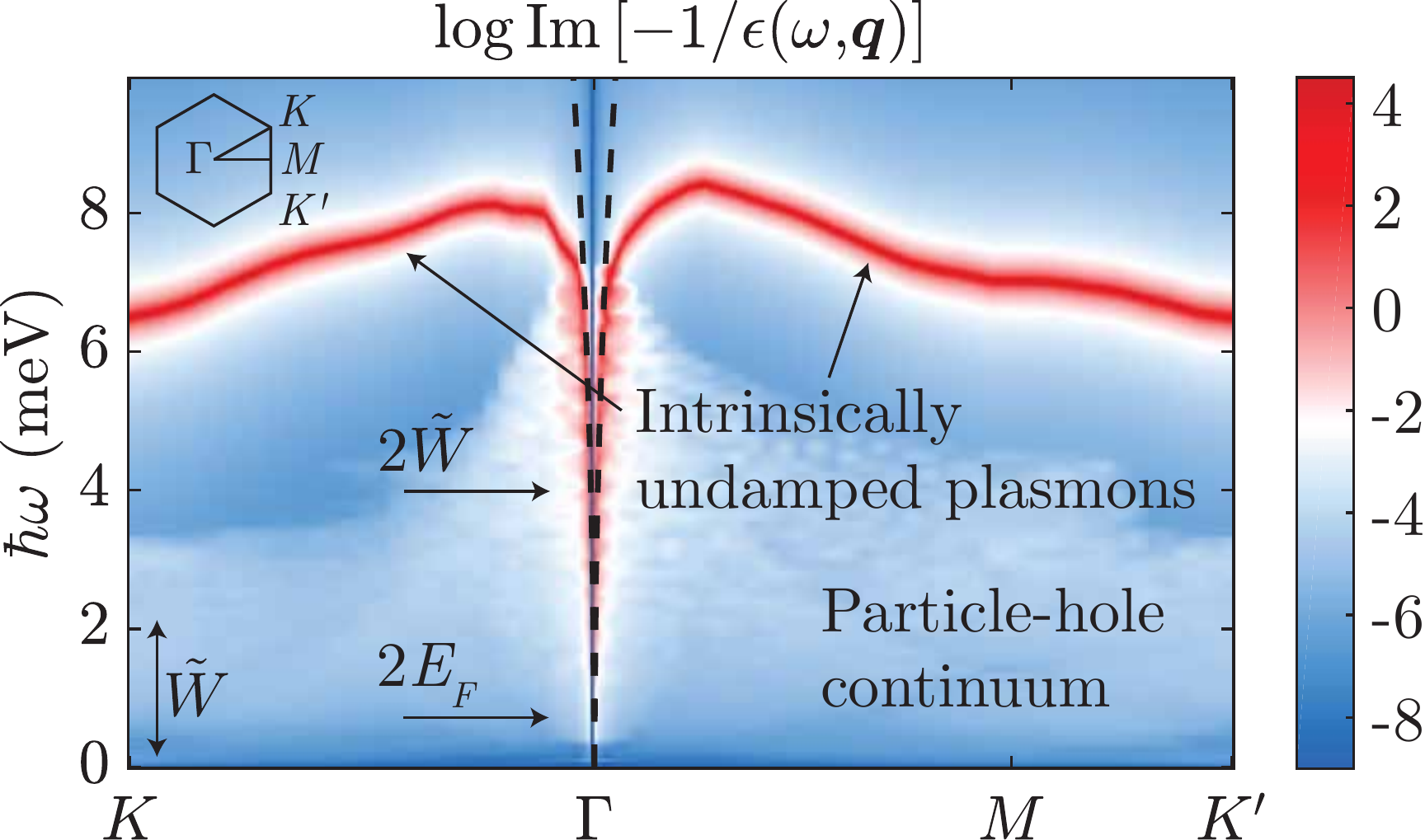}
    \caption{Electron loss function $\Im(-1/\epsilon(\omega,\vec{q}))$ for TBG bandstructure. 
%We placed 
The Fermi energy value $E_{\rm F} = 0.289$ meV corresponds to electron band half-filling, and %$n=1.33\times10^{12}$ $cm^{-2}$ 
the average background dielectric constant is $\kappa = 3.03$ (typical of an air/TBG/hBN heterostructure).   Log scale is used to clarify the relation between different features. Arrows mark the approximate interband p-h continuum edges 
obtained for the effective bandwidth $\tilde{W} \approx 2$ meV %as discussed in the main 
(see text).
    %The dashed line is a fit to 
    Plasmon dispersion (red line) at small $q$ is fitted with $\omega_{\rm p}(q) = \sqrt{\beta_q q}$ [Eq. \eqref{eq:omega=sqrt(Bq)}]  %,  \eqref{eq:beta_0} 
    %  \eqref{eq:beta_q_powers}
    %Eqs.    \eqref{eq:plasmon_dispersion_tb},  \eqref{eq:plasmon_beta_tb} 
    (dashed line), demonstrating a significant deviation from the typical 2D plasmon dispersion at large $q$.  In the calculation, we used both flat bands and the next conduction/valence nonflat bands and verified that higher bands do not alter the quantitative and qualitative behavior. % result drastically. %(c) Small $q$ dependence of the plasmon dispersion. Function $B(\vec{q})$ is given by   \eqref{eq:B_coeff} with the summation over bands limited to the flat-bands only. The dashed curve is given by $\omega_{\rm p} = \beta_0 q + \beta_1 q^2$, where $\beta_0$ and $\beta_1$ are fitting parameters.% is well fitted by $\omega_{\rm p}^2 \approx \beta_0 q$ with $\beta_0 \approx 5459$ $meV^{2} ~nm$.\addCL{[CL: I could also fit $\omega_{\rm p}^2 = \beta_0 q + \beta_1 q^2$?]}
}
    \label{fig:fig_3}
  \vspace{-5mm}
\end{figure}

%\addCL{Dielectric function of TBG}
The definition of the polarization function for the TBG continuum model is essentially identical to that of the tight binding toy model [Eq. \eqref{eq:pol_tight_binding}]. Now, however, we must account explicitly for  the valley and spin degrees of freedom, for a larger number of electron bands, and for different coherence factors. Accordingly, we promote the band indices $s$,$s'$ in Eq. \eqref{eq:pol_tight_binding} to composite labels $n$,$m$, which label all electron bands, spins $\sigma$  and valleys $\xi$; this makes the additional factor of $4$ in front of Eq. \eqref{eq:pol_tight_binding} redundant. %The last difference between 
The toy model coherence factors are replaced by the TBG coherence band factors $F^{nm}_{\vec{k}+\vec{q},\vec{k}}$, which are given by
\be
F^{nm}_{\vec{k}+\vec{q},\vec{k}} = \left\lvert \int_{\Omega} d^2 r \Psi^\dagger_{n,\vec{k}+\vec{q}}(\vec{r}) e^{i \vec{q} \cdot \vec{r}} \Psi_{m,\vec{k}}(\vec{r}) \right\rvert^2
,
\label{eq:tbg_coherence_factor}
\ee
where $\Psi_{n,\vec{k}}(\vec{r})$ are the Bloch wavefunctions for momentum $\vec{k}$ and band/valley/spin composite label $n$, which diagonalize the continuum Hamiltonian (see supplemental materials). The integral in Eq. \eqref{eq:tbg_coherence_factor} is carried over the moir{\'e} unit cell $\Omega$.

%\addCL{TBG plasmons - numerical solution}
After the polarization function is evaluated, we can %straightforwardly 
determine the dielectric function and identify TBG's collective modes from poles of $1/\epsilon(\omega,\vec q)$ as above. An example of a TBG's dielectric function at approximately half-filling of the electron band, $E_{\rm F} = 0.289$ meV, is shown in Fig.  \ref{fig:fig_3}; fixed $q$ line cuts and zeros of $\epsilon(\omega,\vec q)$ are illustrated in the supplemental materials. In discussing the figure, it is helpful to contrast it with the calculation for the hexagonal-lattice toy model shown in Fig.   \ref{fig:fig_1}a. We again see a well-defined intrinsically undamped plasmon mode $\omega_{\rm p}$ (red in Fig.   \ref{fig:fig_1}a) positioned above the p-h continuum; the mode resides inside the band gap $2W < \omega_{\rm p} < W+\Delta$, which peaks at $\hbar \omega_{\rm p} \approx 8.5$ meV before decreasing and becoming almost flat $\hbar \omega_{\rm p} \approx 6.5$ meV at large momenta. %As expected from earlier
In agreement with the analytic considerations above, we see the interband continuum extending from $2 E_{\rm F}$  to $2 W$, but since $E_{\rm F} = 0.289$ meV is extremely small, it makes the conventional (Landau-damped) part of plasmon dispersion $\omega < 2E_{\rm F}$ invisible on the figure. 

%\addCL{TBG plasmons - analytical considerations} 
There are several unique aspects of the TBG plasmon dispersion compared with the behavior of generic narrow-band plasmons discussed above. To analyze the dispersion at $\omega_{\rm p}>2W$, we proceed just as in the toy model case, rewriting the TBG polarization function in a slightly different form of Eq. \eqref{eq:dielectric_function_tb_step_2}, where the indices $n$, $m$ and the band coherence factor are modified as described %in the paragraph 
above.

To proceed further analytically,  we need to analyze Eq. \eqref{eq:dielectric_function_tb_step_2} in the long-wavelength limit. However, unlike the 2-band toy model, where the only characteristic energy scale was the bandwidth $W$, the TBG band structure features an additional energy scale, namely, the gap between the flat bands and the rest of the energy spectrum. This impacts the small-$q$ series expansion of the polarization function, as now the energy difference $E_{n} - E_{m}$ between the occupied and unoccupied states can be larger than $\omega$. To account for such contributions in the series expansion, we split the summation over TBG bands into 2 parts, depending on whether $\omega$ or the energy difference $E_{n} - E_{m}$ is the largest energy scale in the denominator of Eq. \eqref{eq:dielectric_function_tb_step_2}. This yields an approximate expression for the dielectric function
\be
\epsilon(\omega,\vec{q}) \approx 1 + A(\vec{q})-\frac{B(\vec{q})}{\omega^2}
,
\label{eq:die_simp}
\ee
where we defined 2 auxiliary functions:
\be
    A(\vec{q}) = \frac{8 \pi e^2}{\kappa q} \sideset{ }{'}\sum_{\vec{k},n,m} f_{m,\vec{k}} \frac{F^{n m}_{\vec{k}+\vec{q},\vec{k}}}{E_{n,\vec{k}+\vec{q}}-E_{m,\vec{k}}}\,
   % \frac{\left\lvert\left\langle \psi_{\xi,m,\vec{k}+\vec{q}} | e^{i\vec{q}\cdot \vec{r}} | \psi_{\xi,n,\vec{k}}\right\rangle\right\rvert^2}{\epsilon_{\xi,m,\vec{k}+\vec{q}}-\epsilon_{\xi,n,\vec{k}}}\,
\ee
and
\be
    B(\vec{q}) = \frac{8 \pi e^2}{\kappa q} \sideset{ }{''}\sum_{\vec{k},n,m} f_{m,\vec{k}} F^{nm}_{\vec{k}+\vec{q},\vec{k}}(E_{n,\vec{k}+\vec{q}}-E_{m,\vec{k}})\label{eq:B_coeff}
.
\ee
Here the band summations $\sum^{\prime}$ and $\sum^{\prime\prime}$ run over bands such that $\omega^2 > (E_{n,\vec{k}+\vec{q}}-E_{m,\vec{k}})^2$ and $\omega^2 < (E_{n,\vec{k}+\vec{q}}-E_{m,\vec{k}})^2$, % hold 
respectively: for example, at large momenta, as seen in Fig.   \ref{fig:fig_3}, the plasmon mode lies in the gap between the flat and non-flat bands, and hence, the $B(\vec{q})$ summation extends only over the flat bands, whereas the summation in $A(\vec{q})$ includes all of the remaining combinations of band indices. This allows us to write a closed form expression for the plasmon dispersion as
\be
\omega_{\rm p}^2 \approx \frac{B(\vec{q})}{1+A(\vec{q})}
,
\label{eq:plasmon_dispersion_tbg}
\ee
which %we expect to 
must hold for both small and large $q$. We consider these 2 limits separately. 

At small $q$, the matrix element of the Bloch wavefunctions, just as in the toy model case, favors the overlap between states from the same band. At the same time, there are fewer states in the $A(\vec{q})$ satisfying the condition $\omega^2 > (E_{n,\vec{k}+\vec{q}}-E_{n,\vec{k}})^2$, and hence, $A(\vec{q})$ vanishes for small $q$. This amounts to the plasmon dispersion $\omega_{\rm p}$ from Eq. \eqref{eq:plasmon_dispersion_tbg} reducing to $\omega_{\rm p}^2 \approx B(\vec{q})$, and by comparison with Eq. \eqref{eq:intraband_pol_approximation}, we similarly expect a conventional 2D plasmon dispersion $\omega_{\rm p} = \sqrt{\beta_q q}$ with $\beta_q$ given by the series from Eq. \eqref{eq:beta_q_powers}. As we see in Fig. \ref{fig:fig_3}, the $\omega_{\rm p} = \sqrt{\beta_q q}$ dispersion is a valid description only at very small $q$ compared to the Fig. \ref{fig:fig_1}a, which can be traced back to higher  bands softening the plasmon dispersion through the $A(\vec{q})$ term in Eq. \eqref{eq:plasmon_dispersion_tbg}.

%In the other limit of $q$ becoming 

To determine how high the plasmon mode rises above the p-h continuum, we consider large $q$ values comparable to the reciprocal lattice vector. The arguments similar to those in the toy model show that, since $\alpha\gg 1$, we have $A(\vec{q})\gg 1$. The dependence on the $e^2/\kappa q$ ratio, therefore, cancels between the $A(\vec{q})$ and $B(\vec{q})$ functions, resulting in the value of the plasmon dispersion $\hbar \omega_{\rm p} \approx \sqrt{B(\vec{q})/A(\vec{q})} \sim \sqrt{W \Delta} \approx 6.6$ meV being dictated only by the continuum model's band structure parameters. % and, most crucially, larger then the continuum located at energies $\sim\!W$. 
This lack of explicit dependence on $\alpha$ suggests that, after the doping is such that $\alpha \gg 1$, the large-$q$ value of $\hbar \omega_{\rm p} \approx \sqrt{W \Delta}$ becomes insensitive to %the actual value of 
doping (and hence, Fermi velocity). This behavior is %unseen in 
different from that in the toy model, where $\omega_{\rm p} \sim\sqrt{\alpha} W$ at large $q$. The relatively more weak dependence on $\alpha$ %, and its origin 
in the TBG case is due to interband polarization involving higher bands, which significantly alters the effective dielectric constant. The weak $q$ dependence at large $q$ is in agreement with the properties of interband plasmons described in ref. \cite{stauber2016}.

We also note that, although plasmons above the p-h continuum are kinematically protected from p-h excitation, which makes them undamped at the RPA level, there exist relaxation pathways through higher-order pair production in which several electron-hole pairs are emitted with total energy exceeding $\tilde{W}$, as well as phonon-assisted processes. For conventional plasmons these processes were analyzed in ref. \cite{glazman_damping}. The role of these effects for plasmon lifetimes in TBG will be a subject of future work.

%\addCL{Outlook}
Before closing, we note that suppressing damping has always been central to the quest for tightly-confined low-loss surface plasmon excitations. An early approach utilized surface electro-magnetic modes traveling at the edge of an air/metal boundary  \cite{Raether}, in which dissipation is low because most of the mode field resides outside the metal; however, the field confinement scale, set by optical wavelength, was fairly large. Next came surface plasmons propagating in high-mobility 2D electron gases in semiconductor quantum wells and monolayer graphene  \cite{woessner2014}, which can provide deep-subwavelength confinement \cite{jablan}. However, plasmons in these systems are prone to a variety of dissipation mechanisms, with Landau damping usually regarded as the one that sets the fundamental limit on possible plasmon wavelengths and corresponding lifetimes. The possibility to overcome this fundamental limitation in narrow-band systems, such as moir\'e graphene, creates a unique opportunity for graphene plasmonics. 
Damping-free plasmons can enable novel interference phenomena, dissipationless photon-matter coupling, and other interesting behaviors.
%pave way to interesting fundamental behaviors such as novel interference phenomena, dissipationless photon-matter coupling, and 
It is also widely expected that low-dissipation plasmons can lead to unique applications for photon-based quantum information processing  \cite{gullans2013}. 
Furthermore, reduced damping has %a range of 
more immediate consequences, as it translates into enhanced optical coherence that can be directly probed by scanning near-field microscopy, as discussed above, providing a clear signature of the undamped collective modes.

%The search for highly confined low loss plasmons has a long-standing history in the field of condensed matter starting with surface plasmons travelling at the edge of an air/metal boundary \cite{Raether} and ending with surface plasmons propagating in graphene-h-BN heterostructures \cite{woessner2014}. Although surface plasmons in two dimensional heterostructures are perhaps the only viable strategy for confining and controlling light at a nanoscale \cite{jablan}, the range of possible plasmon wavelengths and corresponding lifetimes is ultimately tied with Landau damping. In this article we demonstrated how surface plasmons in narrow band systems - characterized by a high effective fine structure constant $\alpha \gg 1$ - overcome this fundamental limitation and can exist in a regime with a completely suppressed Landau damping. The degree to which plasmons are undamped is tunable \textit{in situ} by varying Fermi energy, which in turn changes the effective fine structure constant $\alpha$ and hence the extent to which the collective mode rises above the p-h continuum. This gate tunable degree of control allows to induce in a controlled manner an increase in plasmon coherence, which has immediate experimental consequences in scanning near-field optical microscopy measurements and hence forms a clear signature of the intrinsically undamped collective modes.

We thank Ali Fahimniya for useful discussions. This work was supported, in part, by the Science and Technology Center for Integrated Quantum Materials, NSF Grant No. DMR-1231319; and Army Research Office Grant W911NF-18-1-0116.

%\bibliography{references}

	\clearpage

	\setcounter{section}{0}
	\renewcommand{\theequation}{S.\arabic{equation}}
	\setcounter{equation}{0}
    
\section*{Supplemental Information}	
\section{Spatial speckle patterns in near-field optical microscopy} % measurement}

Here we elaborate on the %missing steps between the
analysis connecting Eq. (3) and the %plotted 
speckle patterns shown in Fig. 1d and Fig. 2a-d. For an in-depth discussion of the near-field optical microscopy measurement technique and quantitative modeling of the detected signal we refer the reader to %the supplemental materials of 
refs. \cite{koppens2012, basov2012, woessner2014}.  

As argued in the main text, we can estimate the strength of the measured signal in the near-field opical microscopy by evaluating the equal-point correlation function Eq. (3), which here we restate for convenience:
\be
S(\vec{r}) = J_0 \int d^2 \vec{r}' G_E(\vec{r}-\vec{r'}) \eta(\vec{r'}) G_E(\vec{r}'-\vec{r})
.
\label{eq:sup_correlation_function}
\ee
It describes an amplitude of plasmon excitation, which traveled from the tip at position $\vec{r}$ to a disorder at position $\vec{r'}$ and was then reflected back towards the tip at $\vec r$. Here the Green's function $G_E(r) $ of the plasmon excitation of wavenumber $q_0$ is taken in the limit $rq_0\gg1$ as
\be
G_E(\vec{r}) \approx \frac{e^{i q_0|\vec{r}|}}{\sqrt{2\pi |\vec{r}|}}e^{-\delta|\vec{r}|}
, \label{eq:plasmon_green_function}
\ee
which %is simply the Green's function of a 
describes radially propagating waves in 2D. The factor $e^{-\delta|\vec{r}|}$ describes damping due to  extrinsic effects such as phonons and other inelastic processes. Upon substitution of the Green's function into   \eqref{eq:sup_correlation_function}, the measured signal $S(\vec{r})$ is given by
\be
 S(\vec{r}) = J_0 \int d^2 \vec{r}' \eta(\vec{r}')  \frac{e^{i 2 q_0 |\vec{r}-\vec{r}'|}e^{-2\delta|\vec{r}-\vec{r}'|}}{2\pi |\vec{r}-\vec{r}'|}
.
\ee
This expression, which is a convolution of two functions, will generate a product under Fourier transform. 
%which is nothing else than a convolution of two functions. 

For purposes of Fig. 1d and Fig. 2a-d we evaluate the above convolution numerically by using the convolution theorem, that is first performing a fast Fourier transform of both terms individually, multiplying them and then carrying out an inverse Fourier transform. The inset of Fig. 1d and Fig. 2a-d is the intermediate step of this process, but we can also determine it analytically by evaluating the Fourier transform of the signal $S(\vec{r})$
\be
S_\vec{k} = \int d^2 \vec{r} e^{-i\vec{k}\cdot \vec{r}} S(\vec{r})
.
\ee
As expected by the convolution theorem the expression factorizes into a product of two separate factors
\be
S_\vec{k} = \int d^2 \vec{r}' \eta(\vec{r'}) e^{-i\vec{k}\cdot \vec{r}'} \int d^2 \vec{r} \frac{e^{-i\vec{k}\cdot \vec{r}+i2q_0|\vec{r}|}e^{-2\delta|\vec{r}|}}{2\pi |\vec{r}|}
,\label{eq:signal_ft}
\ee
where the first factor is %the definition of 
nothing but the Fourier harmonic of $\eta(\vec{x})$ and the second factor is the $r$-$r'$ influence function, simplified by performing a variable change $\vec{r}-\vec{r}'\to \vec{r}$. To evaluate the %latter 
integral over $d^2\vec r$ we first integrate over $|\vec r|$ and then carry out angular integration using the identity 
% use a standard integral
\be
\int_{0}^{2\pi} d\theta \frac{1}{a+b\cos\theta} = \frac{2\pi}{\sqrt{a^2-b^2}}
.
\ee
After substituting $a=k$, $b=2k_0+2\delta i$ this gives Eq. (4) of the main text.

\section{Behavior of the %toy-model's 
intraband and interband polarization functions}

Here we discuss the behavior of the intraband and interband polarization functions $\Pi_1(\omega,\vec q)$ and $\Pi_2(\omega,\vec q)$ of the toy model, defined in Eqs.  (12), (13) of the main text. 
In particular,  we estimate the coefficients
 $\lambda_1$ and $\lambda_2$ describing the small-$q$ behavior of $\Pi_1$ and $\Pi_2$, defined in the paragraph beneath Eq.(14). We are mostly interested in high frequency values $\omega >2W$ describing the intrinsically undamped regime.

%the polarization function, as discussed in the main text for the toy model. 

We start with the quantity $\lambda_2$ describing the contribution of intraband transitions. At small $q$, the interband coherence factor from Eq. (9)
is non-negligible only in proximity of the points $K$ and $K'$. Near these points a linear dispersion $E_{s,\vec{k}}=s v_F k$, with $s=\pm 1$, is a good approximation for the bandstructure. In that limit, the small-$q$ interband coherence band factor from Eq. (11) becomes
\be
F^{s=- s'}_{\vec{k}+\vec{q},\vec{k}} \approx \frac{1}{4}\lp\vec q\cdot\nabla_{\vec k}\phi_{\vec k}\rp^2 \approx \frac{1}{4} \frac{q^2}{k^2} \sin^2 \theta
,
\label{eq:coherence_factor_approximation}
\ee
where $\theta$ is the angle between $\vec{k}$ and $\vec{q}$. The quantity $\Pi_2(\omega,\vec{q})$ is therefore given by:
\be
\Pi_2(\omega,\vec{q})=-\frac{8 q^2}{\omega^2} \sum_{\vec{k},s} f_{s,\vec{k}} s v_F k \frac{\sin^2{\theta}}{k^2}
.
\ee
In the above we used the linear dispersion approximation $E_{s,\vec{k}} = s v_F k$ for the whole band and accounted for the $K$ and $K'$ points through an additional factor of $2$. This gives
\be
\Pi_2(\omega,\vec{q}) \approx -\frac{2 E_F}{\pi} \frac{q^2}{\omega^2}+\frac{2 W}{\pi} \frac{q^2}{\omega^2} = \frac{2}{\pi} (W-E_F) \frac{q^2}{\omega^2}
,
\ee
with the first and second terms originating from the conduction band and the valence band respectively. This gives
\be
\lambda_2 = 2 (W-E_F)/\pi
,
\ee
which takes positive values since $-W<E_F<W$. 

Next, we proceed to estimate the $\lambda_1$. Without loss of generality, we place the Fermi energy in the conduction band. In this case,  the interband contribution to the polarization function is non-vanishing only in the conduction band. This can be seen by going back to the Eq. (12), which for $s=s'=-1$ and small $q$ vanishes:
\begin{align}
\Pi_1(\omega,\vec q)&\approx-\frac{8}{\omega^2} \sum_{\vec{k}} f_{-1,\vec k} (E_{-1,\vec{k}}-E_{-1,\vec{k}+\vec{q}}) \\
&= -\frac{8}{\omega^2} \sum_{\vec{k}} (E_{-1,\vec{k}}-E_{-1,\vec{k}+\vec{q}}) = 0
,
\end{align}
since $f_{-1,\vec{k}}=1$ for all $\vec{k}$ in the valence band. It is therefore sufficient to focus on the contribution of the partially filled (conduction) band. To be consistent with the $\lambda_2$ analysis above we replace the dispersion energy as $E_{1, \vec{k}} = v_F k$.  The intraband contribution to the polarization function $\Pi_1(\omega,\vec{q})$ is then
\be
\Pi_1(\omega,\vec{q}) \approx \frac{8 q^2}{\omega^2} \sum_{\vec{k}} f_{1,\vec{k}} v_F \frac{\sin^2\theta}{k} = \frac{2}{\pi} E_F \frac{q^2}{\omega^2},
\label{eq:interband_small_q_evaluated}
\ee
giving 
\be
\lambda_1 = 2 E_F / \pi
.
\ee 
As argued in the main text [see discussion below Eq. (14)], this result remains unchanged for frequencies $\omega < 2 E_F$ and, therefore,
\be
\beta_0 = 4\alpha v_F E_F
.
\label{eq:beta_0}
\ee
Going back to the $\omega > 2 W$ regime, and using $\lambda_1$ and $\lambda_2$ derived above, gives the square-root plasmon dispersion $\omega_{\rm p} = \sqrt{\beta q}$ with
\be
\beta=\frac{2\pi e^2}{\kappa}(\lambda_1+\lambda_2) = 4 \alpha v_F W
.
\ee
Therefore, at small $\omega < 2 E_F$ the dispersion behaves as $\omega_{\rm p} = 2 \sqrt{\alpha v_F E_F q}$, becoming enhanced at high energies $\omega > 2 W$, $\omega_{\rm p} = 2 \sqrt{\alpha v_F W q}$ by a factor $\sqrt{W/E_F}$.

To complete the analysis of the polarization function behavior, now we focus on  frequencies in the region $2 E_F< \omega < 2 W$. Working again in the small-$q$ limit we find that, as pointed out earlier, only the interband contribution to the polarization function $\Pi_2(\omega,\vec{q})$ develops an imaginary part, whereas the intraband polarization function $\Pi_1(\omega,\vec{q})$ is real-valued, %simply 
given by the   Eq. \eqref{eq:interband_small_q_evaluated}. To determine the form of $\Pi_2(\omega,\vec{q})$ in the interband p-h continuum energy range, we approximate the coherence factor as in   Eq. \eqref{eq:coherence_factor_approximation} to obtain
\be
\Pi_2(\omega,\vec q) \approx 8\sum_{\vec{k},s} f_{s,\vec k}  \frac{s v_F k}{4v_F^2 k^2-(\omega+i0)^2}\times  \frac{q^2}{k^2} \sin^2 \theta
.
\ee
Here we used the linear approximation to the energy dispersion $E_{s,\vec{k}} = s v_F k$, accounting for the fact that, because of the behavior of the coherence factors, only the states near the Dirac point contribute  to $\Pi_2$. As always, we account for the $K$ and $K'$ points by an additional factor of $2$. After carrying out  integration over $d^2k$ we arrive at:
\be
\Pi_2(\omega,\vec q) \approx -i\frac{2 q^2}{\omega}\Theta(\omega-2E_F)\Theta(\omega-2W)
.
\ee
Here $\Theta(x)$ is the Heaviside function, which ensures that the imaginary part is non-zero only in the particle-hole continuum region $2 E_F < \omega < 2 W$. The dielectric function in this region is therefore
\be
\epsilon(\omega,\vec{q}) = 1 - \beta_0\frac{q}{\omega^2} + i \frac{\beta_0 \pi}{E_F} \frac{q}{\omega}
,
\ee
which shows that the collective mode $\omega_{\rm p}$ in the $2 E_F < \omega_{\rm p} < 2 W$ region has a damped square-root dispersion
\begin{align}
%\omega_{\rm p} &= \sqrt{\beta_0 q - \frac{\pi^2 \beta_0^2}{4 E_F} q^2}-i \frac{\pi \beta_0}{2E_F} q\nn
\omega_{\rm p} &\approx \sqrt{\beta_0 q}-i \frac{\pi \beta_0}{2E_F} q
.
\end{align}
%is a $\sqrt{\beta_0 q}$ dispersing plasmon, which gets rapidly damped by pair production.
The imaginary part, which scales linearly with $q$, describes damping due to particle-hole pair production.

We finish the discussion of the collective modes %in the toy model 
by comparing the analytically predicted dispersion with the numerical result  in Fig. 1a. While the simulated dispersion closely follows the square-root dependence $\omega_{\rm p} \propto \sqrt{q}$, %relation well, we achieve an almost perfect 
the agreement between the simulation and $\omega_{\rm p} = \sqrt{\beta_q q}$ dispersion %becomes nearly perfect 
is drastically improved if the two first terms $\beta_0$ and $\beta_1$ from the series expansion in Eq. (14) are used for a fitting. Although the terms $\beta_0$ and $\beta_1$ could in principle be computed by carrying out an expansion of the polarization function %from 
in Eq. (10) in powers of $q$ and then evaluating the resulting integrals numerically, we instead treat $\beta_0$ and $\beta_1$ as free parameters and %instead 
fit them to the simulated dispersion. This approach yields values
\begin{align}
\label{eq:beta_0_beta_1_est}
\beta_0 &= 0.96\times10^3 ~{\rm meV}^2 ~{\rm nm}
,\nn
 \beta_1 &= -10^3 ~{\rm meV}^2 ~{\rm nm^2}
.
\end{align}
%where 
The best-fit $\beta_0$ value is close to $\beta_0 = 4\alpha v_F E_F \approx 0.86 \times10^3$ ${\rm meV}^2 ~{\rm nm}$ %is expected 
predicted from   Eq. \eqref{eq:beta_0}. We also see that, since $\beta_1$ is negative, % indeed $\beta_1 < 0$ softening 
the plasmon dispersion is indeed softened by interband polarization, in agreement with the argument given in the main text [see Eq. (14)]. 

\section{Twisted bilayer graphene - details of the model}
Here we describe in detail the model for twisted bilayer graphene (TBG) bandstructure used in the main text. %and its eigenstates with the help of the
We use the effective continuum Hamiltonian introduced in ref. \cite{koshino2018}, adopting notations and numerical values used in ref. \cite{koshino2018}. 

The continuum approach is made possible by the small values of the twist angle $\theta$ by which the two graphene layers in TBG are rotated relative to one another. 
%For clarity of the analysis let us discuss the process of constructing the continuum Hamiltonian. 
We start by taking two AA-stacked graphene layers and rotating the layer 1 and the layer 2 around the B-sites by $-\theta/2$ and $\theta/2$ respectively. %The angle $\theta$ we set at 
For the ``magic'' value of $\theta = 1.05^{\circ}$, %which makes 
the moir\'e real-space lattice constant is $L_M = a/2\sin(\theta/2) \approx 13.4$ nm. This is  two orders of magnitudes greater than the graphene's lattice constant $a = 0.246$ nm, justifying the use of the continuum approach.

In momentum space this real-space rotation translates into two graphene Brillioun zones rotated by angle $\theta$ relative to each other. Both BZs are centered at the same $\Gamma$ point but the $K$ (and $K'$) points of the two layers are separated by a small momentum $4\pi/(3 L_M)$. As the moir{\'e} periodicity $L_M$ is much greater than the lattice constant $a$, we can ignore the intervalley mixing between the two valleys $K$ and $K'$ of the original graphene layers - here labeled by $\xi=-1,1$. The total Hamiltonian of the system becomes therefore block diagonal in the valley index. The blocks $H^{(\xi)}$ describing each of the two valleys take the form
\be
H^{(\xi)}=\begin{pmatrix} 
H_1 & U^\dagger \\
U & H_2 
\end{pmatrix}\label{eq:tbg_ham}
\ee
in the basis of $(A_1, B_1, A_2, B_2)$ sites. The matrices $H_{l}$ ($l=1,2$) correspond to the intralayer Hamiltonians of the layers. The latter, due to the lengthscale separation between $L_M$ and $a$, can be approximated by performing the standard $kp$ expansion around the points $K$ and $K'$. 

This procedure gives $2\times 2$ Dirac Hamiltonians centered at the $\vec{K}_{\xi}^{(l)}$ points
\be
H_{l}= -\hbar v \left[ R\left(\pm\theta/2\right)(\vec{k}-\vec{K}_{\xi}^{(l)})\right] \cdot \left(\xi\sigma_x,\sigma_y\right)
,
\label{eq:linear_equation}
\ee
where $\vec{k}$ is a momentum in the BZ of the original graphene layers, and  $R\left(\phi\right)$ is the $2\times 2$ %2D counterclockwise 
rotation matrix
\begin{align}
R\left(\phi\right) = \begin{pmatrix} 
\cos\phi & -\sin\phi\\
\sin\phi & \cos\phi 
\end{pmatrix}
\end{align}
that accounts for rotation of the BZ of the original graphene layers.  The signs $\pm$ in   Eq. \eqref{eq:linear_equation} correspond to the layers $l = 1$ and $2$, respectively. 

The energy scale for the Hamiltonians $H_l$ is $\hbar v/a = 2.1354$ eV. 
The vectors $\vec{K}_{1}^{(l)}$, $\vec{K}_{-1}^{(l)}$, which denote the Dirac points $K$ and $K'$ of the layers, are given by
\be
\vec{K}_{\xi}^{(1)} = -\xi \frac{4\pi}{3a} R\left(-\theta/2\right) \begin{pmatrix} 
1 \\
0
\end{pmatrix},\quad 
\vec{K}_{\xi}^{(2)} = -\xi \frac{4\pi}{3a} R\left(\theta/2\right) \begin{pmatrix} 
1 \\
0
\end{pmatrix}
,
\ee
respectively. We stress that, while $\vec{k}$ alone has length close to $\sim4\pi /3a$, the difference $\vec{k}-\vec{K}_{\xi}^{(l)}$ is small, since $\vec k$  is always located near the vicinity of the $\vec{K}_{\xi}^{(l)}$ points. This makes the linear expansion from   Eq. \eqref{eq:linear_equation} a well defined approximation.

More quantitatively, the expressions in   Eq. \eqref{eq:linear_equation}, found by Taylor expanding the graphene tight-binding Hamiltonian, are valid for momenta close enough to  the Dirac points of the two layers, $|\vec{k}-\vec{K}_{\xi}^{(l)}|a\ll 1$. For $\theta\ll 1$ this condition is obeyed in the entire mini Brillouin zones of the TBG superlattice. 

%
% and are explicitly given as
%\be
%H_{l}= -\frac{\hbar v}{a}\begin{pmatrix} 
%0 &e^{-i\xi \theta/2} k_{-} a + \frac{4 \pi}{3} \\
%e^{i\xi \theta/2} k_{+} a + \frac{4 \pi}{3} & 0
%\end{pmatrix}
%,
%\ee
%where $k_{\pm}=\xi k_x \pm i k_y$ and $k_x$,$k_y$ are crystal momenta of the original graphene layers. Here we take the energy scale as $\hbar v/a = 2.1354$ eV and the lattice constant $a$ as the graphene's lattice constant $a = 0.246$ nm. The angle $\theta$ in this section is the twist angle between the two layers, which we set at the magic angle of $\theta = 1.05^{\circ}$. 

In the analysis below the moir{\'e} superlattice BZ is defined as in the inset of Fig. 3, with the two reciprocal lattice vectors
\be
    \vec{G}_1^{M} = -\frac{2\pi}{\sqrt{3} L_M} \begin{pmatrix} 
1 \\
\sqrt{3} 
\end{pmatrix}
,\quad 
\vec{G}_2^{M} = \frac{4\pi}{\sqrt{3} L_M} \begin{pmatrix} 
1 \\
0 
\end{pmatrix}
.
\ee
We denote the reciprocal lattice vector length as $G_ M = |\vec{G}_1^{M}| = |\vec{G}_2^M| = 4\pi / \sqrt{3} L_M$. Matrix $U$ is the effective moir{\'e} interlayer coupling given by:
\begin{align}
U = \begin{pmatrix}
u & u'\\
u' & u\\\end{pmatrix}&+\begin{pmatrix}
u & u'\nu^{-\xi}\\
u'\nu^{\xi} & u\\\end{pmatrix}e^{i\xi \vec{G}_{1}^M\cdot\vec{r}}+\nn
&+\begin{pmatrix}
u & u'\nu^{\xi}\\
u'\nu^{-\xi} & u\\\end{pmatrix}e^{i\xi \left(\vec{G}_{1}^M+\vec{G}_{2}^M\right)\cdot\vec{r}}
,
\end{align}
where we introduced a notation for the phase factor $\nu=e^{i2\pi/3}$. The interlayer couplings $u$ and $u'$ are taken as $u=0.0797$ eV and $u'=0.0975$ eV to match values in ref. \cite{koshino2018}.

To determine the energy bands and the eigenstates we take the Bloch wavefunction ansatz for a valley $\xi$ as
\be
\Psi_{\xi,n,\vec{k}}^{X}(\vec{r})=\sum_{\vec{G}} C_{\xi,n,\vec{k}}^{X}(\vec{G})e^{i(\vec{k}+\vec{G})\cdot \vec{r}}\label{eq:bloch_ansatz}
\ee
with $X$ %corresponding to each of 
labeling the spinor components $X=A_1$, $B_1$, $A_2$, $B_2$. The band index, labeled by $n$ and $\vec{k}$, is the Bloch wave vector in the BZ of the original graphene layers. Here $\vec{G}$ runs over all possible integer combinations of the reciprocal lattice vectors, $\vec{G}=m_1 \vec{G}_{1}^{M}+ m_2 \vec{G}_{2}^{M}$ with integer $m_1$ and $m_2$. As discussed in ref. \cite{koshino2018}, the low-energy states are expected to be dominated by %the individual graphenes' eigen
states near the original Dirac points. We therefore take only not-too-large indices $m_1$ and $m_2$ that satisfy the condition
\begin{align}
| \vec{k} + \vec{G} - \vec{M_{\xi}} | \le z G_M\,,
\end{align}
where $z$ is a conveniently chosen number of order one [ref. \cite{koshino2018} uses $z=4$], and $\vec{M_\xi}$ are the ``mean'' Dirac point locations
\begin{align}
\vec{M_\xi} = \frac{1}{2}\left(\vec{K}_{\xi}^{(1)} + \vec{K}_{\xi}^{(2)} \right) = -\frac{4\pi}{3 a} \xi \cos(\theta/2) \begin{pmatrix} 
1 \\
0 
\end{pmatrix}
,
\end{align}
given by the midpoint between the $K$ (or $K'$) points of the two layers. 

\section{Electron loss function for the TBG bandstructure}
%\addCL{

\begin{figure}[!htp]
    \centering
    \includegraphics[width=0.99\linewidth]{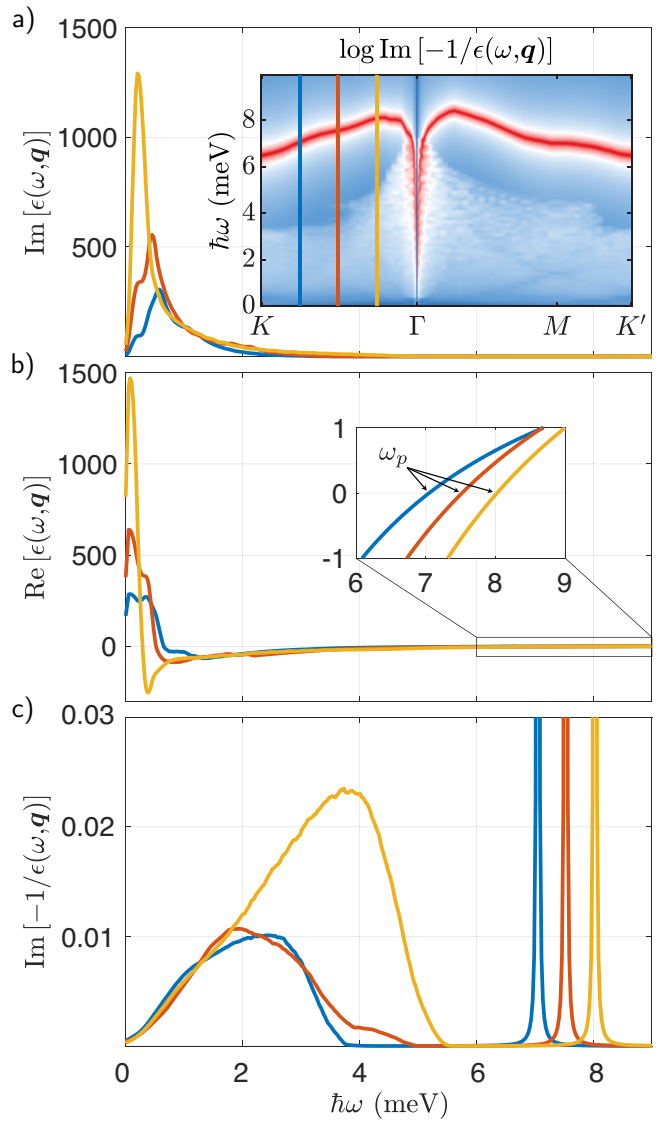}
    \caption{ Imaginary (a) and real (b) parts of the TBG's dielectric function $\epsilon(\omega,\vec{q})$  for several momenta values marked by the colored lines in the inset of (a). The zoom-in in panel (b) shows the positions of plasmon resonances found from $\epsilon(\omega,\vec q)=0$. The inset in (a) is a replica of the loss function shown in Fig.  3; higher-resolution linecuts at the selected momenta are presented in (c). The curves in (a-c) were smoothed with an equal-weighted moving filter.
}
    \label{fig:sup_fig_4} 
  \vspace{-5mm}
\end{figure}

Fig. \ref{fig:sup_fig_4} details the behavior of the electron loss function for TBG, depicted in Fig. 3 of the main text. Panels a and b %of Fig. \ref{fig:sup_fig_4} 
show constant-momentum $\vec{q}$ linecuts of the real and imaginary parts of the dielectric function $\epsilon(\omega,\vec{q})$. The finite width of the plasmon resonance in the loss function in Fig. \ref{fig:sup_fig_4}c is due to the infinitesimal imaginary part of $\omega+i0$ in the polarization function in  Eq. (8) replaced with $\omega+i\gamma$, with a suitably chosen small $\gamma$ introduced for illustration purposes. %}

Strong electron-electron interactions in the narrow electron bands lead to large dielectric function values, as can be seen in Fig. \ref{fig:sup_fig_4}. For energies $\hbar \omega < 2W$ the dielectric function imaginary and real parts take values a few orders of magnitude higher than those of graphene monolayer. The origin of these large values can be traced to the high effective fine structure constant (or, equivalently, low Fermi velocity) in the flat electron bands, as discussed in the main text. To see this in more detail, we recall the  Thomas-Fermi expression for the long-wavelength static dielectric function of graphene \cite{hwang}
\begin{align}\label{eq:TF_epsilon}
\epsilon(\omega = 0,\vec{q} \to 0) = 1 + q_{TF} / q
\end{align}
with the Thomas-Fermi momentum $q_{TF} = N\alpha k_F$, where $N$ is the degeneracy factor $N=8$ ($2$ spins, $2$ layers, $2$ valleys). For illustration purposes, taking a fine structure constant $\alpha\sim 30$ and Fermi momentum $k_F\sim{K}$, for the momentum $q \sim K/2$ (red line in the Fig. \ref{fig:sup_fig_4})    Eq. \eqref{eq:TF_epsilon} predicts a dielectric function value $\epsilon \sim 480$, which is in good agreement with the simulation results.  Above $\hbar \omega > 2W$ %magnitude of 
the dielectric function rapidly decreases until $\hbar \omega > 20$ meV where the contributions of higher electron bands start to dominate. % the value of the dielectric function.

At these energies,  plasmon dispersion is strongly affected by the presence of higher electron bands. At small $q$ plasmon dispersion is %dictated 
predominantly due to intraband transitions, and is thus insensitive to other electron bands. At large $q$ the situation changes. In the absence of higher electron bands the zeros of the dielectric function would occur at much larger energy scales $\hbar \omega_{\rm p} \sim 40 $ meV.  However, as argued in Eq. 19 in the main text, higher electron bands push plasmon dispersion down with the large-$q$ zeros of the dielectric function on the order $\hbar \omega_{\rm p} \approx \sqrt{W \Delta} \approx 6.6 $ meV.  Here $W$ is the flat-band bandwidth and $\Delta$ is the band gap as defined in the main text. The independence of this value of $\alpha$ is the behavior to be expected for large enough $\alpha$,  such that plasmon dispersion extends above the p-h continuum. The independence of $\omega_{\rm p}$ of $\alpha$ at $\alpha\gg 1$  is a characteristic feature of interband plasmons.

Lastly, we note that our simulation is expected to be accurate only for $\vec{q}$ inside the TBG Brillouin zone. When $\vec{q}$ approaches zone boundary  it is necessary to consider local field effects \cite{adler1962, wiser1963}. Although these effects are often small, they require careful examination and thus will be a subject of a future work.

\clearpage

\end{document}